УДК 577.212.3+004.9
ДУПЛИЙ В.П. [1], ДУПЛИЙ С.А. [2]
[1] *Институт клеточной биологии и генетической инженерии НАН Украины,
Украина, 03143, г. Киев, ул. Акад. Заболотного, 148, e-mail: duplijv@icbge.org.ua*
[2] *Mathematical Institute, University of Muenster,
Germany, 48149, Muenster, Einsteinstrasse, 62, e-mail: duplijs@uni-muenster.de*

## TRIANDER – НОВАЯ ПРОГРАММА ДЛЯ ВИЗУАЛЬНОГО АНАЛИЗА НУКЛЕОТИДНЫХ ПОСЛЕДОВАТЕЛЬНОСТЕЙ

Прогресс в молекулярной биологии, связанный секвенированием генов и геномов живых организмов, привел к лавинообразному росту количества данных о молекуле, определяющей все разнообразие жизни. Возможность определения последовательности ДНК в сочетании со статистическими методами дает чрезвычайно важный инструмент для извлечения скрытой информации о динамике процесса эволюции, особенно после того, как стали доступны полные геномы организмов [1]. Наряду с необходимостью создания программного обеспечения для компьютерного анализа накопленных генетических текстов появилась и необходимость в удобном их представлении для человека. В настоящее время имеется множество систем визуализации генетической информации. Большинство из них не показывает тем или иным способом саму последовательность, а только отображает взаимное расположение генов, регуляторных единиц, кодирующих и некодирующих участков.

С другой стороны, использование способности человека к распознаванию образов, при наличии соответствующих инструментов рендеринга сиквенсов, могло бы, благодаря непосредственному восприятию исследователем данных, помочь быстрее выявлять неожиданные феномены эволюции и даже могло бы изменить стиль работы с такими данными.

Нашей задачей было создание оригинальной программы рендеринга нуклеотидных последовательностей (несколькими различными способами) и предоставление пользователю интерактивных возможностей управления визуализацией (масштабирование, выбор участка последовательности и т.д.) в реальном времени.

**Материалы и методы**

Программа для интерактивной визуализации нуклеотидных последовательностей была создана в свободно распространяемой среде разработки программного обеспечения Lazarus версии 1.2.6 [2], использующей компилятор Free Pascal версии 2.6.4 [3].

Программа «Triander» тестировалась под операционными системами Windows XP 32-bit и Windows 7 64-bit. Исходные тексты и скомпилированный бинарный код программы для работы в Windows свободно доступны по адресу http://icbge.org.ua/ukr/Triander .

**Результаты и обсуждение**

Хорошо известно, что сложность визуального анализа генетических текстов, записанных четырехбуквенным алфавитом, значительно возрастает с увеличением длины текста. Человеческий глаз устроен так, что либо читает каждую букву отдельно, не замечая нуклеотидных паттернов, либо отбрасывает последовательности букв, не встречающиеся в привычных словах [4]. Расположение оснований на нотном стане [4] позволяет после некоторой тренировки легче узнавать отдельные паттерны, однако разрыв паттернов при переносе создает гораздо больше проблем, чем при обычном чтении. К тому же, довольно трудно охватить большую последовательность целиком.

В отличие от вышеупомянутых способов визуализации, мы предлагаем естественное представление нуклеотидных последовательностей в виде кривых, где в соответствие основаниям поставлены разнонаправленные векторы. В дальнейшем такие векторы будем называть нуклеотидными векторами, а кривые, образованные ими, нуклеотидными кривыми.

Введенные в [5] «Н-кривые» дают однозначное представление последовательности в трехмерном пространстве и, возможно, были бы идеальными при анализе их на трехмерных устройствах отображения, однако из-за крайне малого распространения таких устройств приходится работать с двухмерными проекциями кривых, что приводит к частичной потере визуальной информации. Двухмерный вариант данного метода [6] получил широкое распространение и оказался полезным для обнаружения в геномах сайтов инициации репликации [7]. Однако из-за отказа от обязательного смещения по вертикали кривая часто проходит по одним и тем же местам диаграммы. Потери информации при этом могут быть значительными, на отдельных участках части кривой сливаются в пятна (рис. 1а, б).

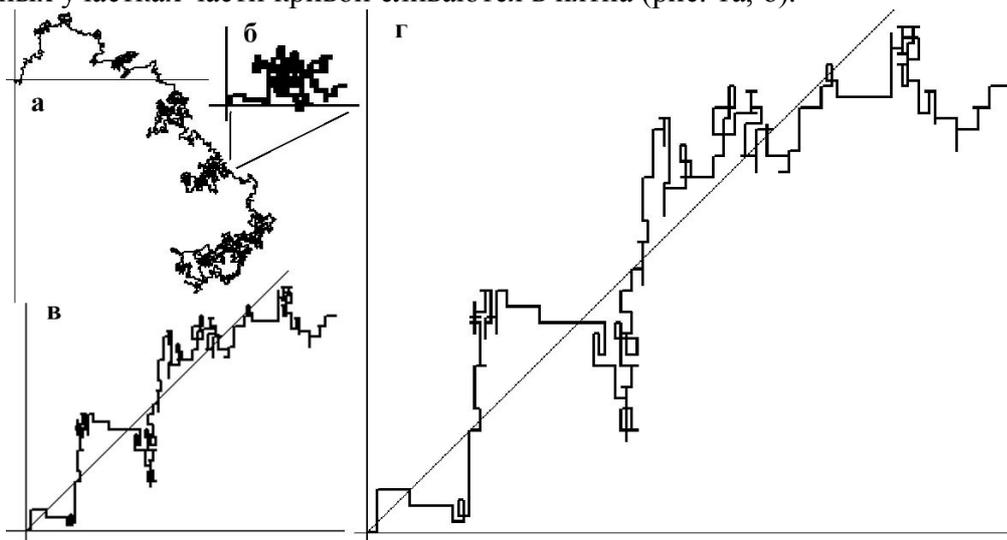

Рис. 1. Диаграммы обхода ДНК последовательности гена нитрат редуктазы Nia1;2 *Physcomitrella patens* (Генбанк AB232049): **а** – полная последовательность; **б-г** – фрагмент AB232049:2601-26300. Для построения использовались нуклеотидные векторы равной длины (**а**, **б**) и пропорциональные степени детерминации нуклеотида (**в**, **г**). Длина единичного вектора: **а-в** – равна ширине нуклеотидной кривой, **г** – больше ширины кривой

В системе визуализации, где нуклеотиды передаются векторами различными не только по направлению, но и по длине [8], в значительной мере эта проблема снимается (рис. 1 в, г). Метод основывается на использовании в качестве длины нуклеотидного вектора его «внутреннюю абстрактную характеристику – степень детерминации» [9]. Степень детерминации – это числовая характеристика нуклеотида, связанная с его способностью определять аминокислоту в зависимости от положения в кодоне, а также с так называемым эволюционным «давлением». Кроме того, принимается во внимание число водородных связей.

Важным является построение именно трех нуклеотидных кривых, которые соответствуют каждому положению в кодоне, что дает построение трех обходов для каждого положения нуклеотида в триплете[10]. При учете степени детерминации такая диаграмма называется триандром [9]. В вышеупомянутой работе также показано, что гипотетическое количество нуклеотидов в кодоне, отличное от трех, а также случайно сгенерированные нуклеотидные последовательности вообще не приводят к появлению таких визуальных структур, как триандры. До настоящего времени не существовало программ для интерактивного построения триандров и диаграмм обхода последовательностей неравными по модулю векторами.

Среди возможных комбинаций направлений нуклеотидних векторов, т.е. векторов, которые представляют тот или иной нуклеотид на диаграмме, нами была выбрана такая: C – Север, G – Юг, T – Восток, A – Запад (рис. 2). Так как степень детерминации, а значит и длина нуклеотидного вектора составляет для C – 4, для G – 3, для T – 2, для A – 1, то диаграмма обхода последовательности случайно выбранных нуклеотидов в нашем случае распространяется в северо-восточном направлении. Это направление, как и четыре основных, указывается на диаграмме.

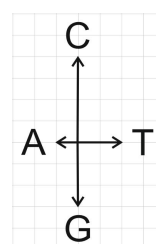

Рис. 2 Направление и длина нуклотидных векторов

Обычно диаграммы обхода ДНК строятся однопиксельными квадратами, поэтому помимо потерь информации из-за прохождения кривой по одним и тем же координатам добавляются потери от слияния лежащих рядом участков кривой. Проблема становится острее при переходе от анализа геномов и хромосом к анализу отдельных генов и регуляторных последовательностей. Мы реализовали возможность задавать длину единичного вектора больше ширины нуклеотидной кривой (рис. 1 г), что с одной стороны сделало диаграммы более читаемыми, а с другой стороны дает возможность таким способом их правильно масштабировать.

Кроме того, большие диаграммы можно смещать по осям координат и задавать для отображения только определенную часть последовательности. Последовательность может быть отображена как в виде триандра (рис. 3 а), ветви которого отображаются кривыми разной толщины, так и обычным методом обхода ДНК, названным в программе по аналогии «монандром». Есть также возможность представить последовательность векторами равной длины. Нужно заметить, что скорости построения диаграмм достаточно для того, чтобы наблюдать анимацию при удерживании кнопки увеличения длины отображаемой последовательности или кнопок смещения ее начала.

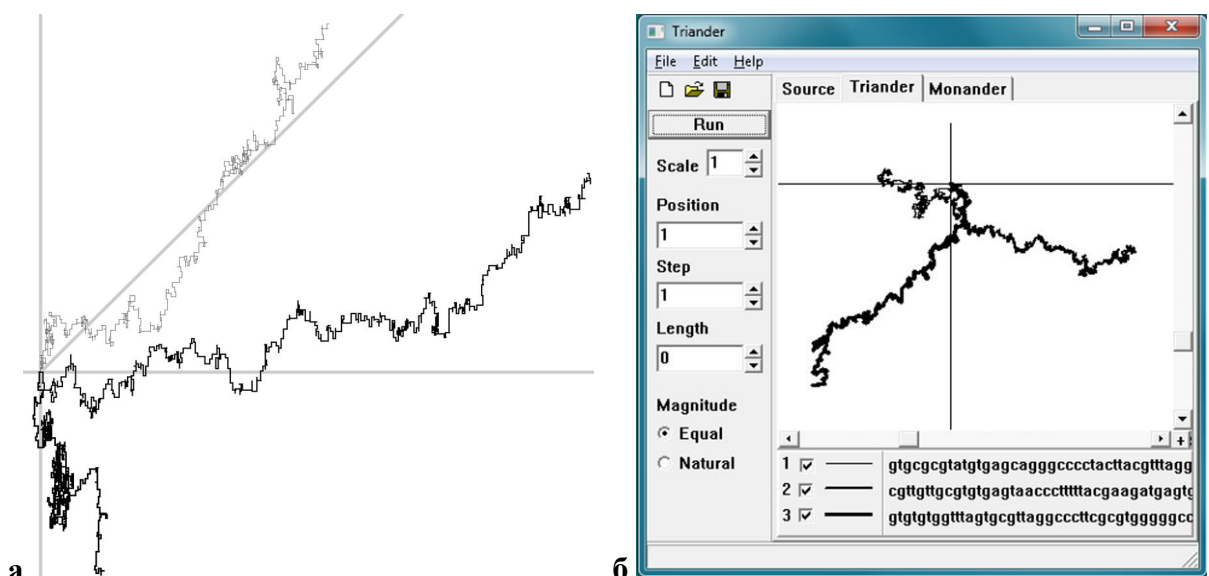

Рис. 3. Диаграммы обхода последовательности DQ157859 в зависимости от положения основания в кодоне. **а** – триандр кодирующей области гена сахарозо-фосфат-синтаза 2 *Physcomitrella patens* (PpSPS2). **б** – главное окно программы «Triander», представляющее обход последовательности равными по модулю нуклеотидными векторами

Созданная нами программа «Triander» (рис. 3б) визуализирует нуклеотидные последовательности, хранящиеся в файлах формата FASTA и GenBank, а также в обычных текстовых файлах. После загрузки файл доступен для просмотра и редактирования. Диаграммы обхода ДНК строятся в широко распространенном формате векторной графики SVG [11], реализована возможность сохранения диаграмм в этом формате.

Наибольшую популярность среди методов графического представления ДНК получили двухмерные диаграммы, построенные обходом последовательности векторами равной длины. Для их построения можно использовать как отдельные программы [12, 13], так и встроенные возможности более крупных проектов [14]. Такие диаграммы хорошо передают структуру больших последовательностей, например хромосом или геномов микроорганизмов. Однако потери визуальной информации из-за наложения частей нуклеотидной кривой друг на друга препятствует эффективному анализу на нуклеотидном уровне.

На сегодняшний день наша программа единственная способна строить триандры и диаграммы обхода ДНК неравными по длине нуклеотидными векторами. Это позволяет получить как общее представление о последовательности, так и различать отдельные паттерны.

**Выводы**

Разработанная нами программа «Triander» позволяет строить несколько вариантов диаграмм понуклеотидного обхода ДНК. Применение внутренней абстрактной характеристики основания, называемой степенью детерминации, в качестве длины нуклеотидного вектора позволяет проводить как общий визуальный анализ на уровне хромосом и геномов, так и выявлять отдельные нуклеотидные паттерны.

**Литература**

**DUPLIJ V.P. [1], DUPLIJ S.A. [2]**
[1] *Institute of Cell Biology and Genetic Engineering of Natl. Acad. Sci. of Ukraine, Ukraine, 03143, Kyiv, Acad. Zabolotnoho str., 148, e-mail: duplijv@icbge.org.ua*
[2] *Mathematical Institute, University of Muenster,
Germany, 48149, Muenster, Einsteinstrasse, 62, e-mail: duplijs@uni-muenster.de*

**TRIANDER – A NEW PROGRAM FOR THE VISUAL ANALYSIS OF THE NUCLEOTIDE SEQUENCE**

*Aims*. Our project aimed to work out the interactive software for nucleotide sequence visualization. *Methods*. The program named as "Triander" was worked out under Free Pascal RAD IDE Lazarus. Source code and compiled for Windows binaries are freely accessible at http://icbge.org.ua/ukr/Triander. *Results*. This program can produce four types of plots. It is possible to build three DNA walks done independently for each nucleotide position in triplets. The usage of not equal in modulus nucleotide vectors lead to significant reduction of visual information loss in DNA walks. *Conclusions*. The program can be used in the investigation of fine structure of sequences and find in them standard patterns and nontrivial regions for further detail analysis.
*Keywords*: DNA walk, triander, determinative degree, software.